\documentclass[prl,twocolumn,amsmath,amssymb]{revtex4}

\usepackage{dcolumn}% Align table columns on decimal point
\usepackage{bm}% bold math
\usepackage{epsfig}
\usepackage[]{graphicx}
\usepackage[]{amsfonts}

\begin{document}

\title{Epitaxial aluminium-nitride tunnel barriers grown by nitridation with a plasma source}
\pagestyle{empty}   %From the page where you put this command there will be no pagenumbering
\pagenumbering{gobble} %No page numbering in the whole document.

\author{T. Zijlstra, C.~F.~J. Lodewijk, N. Vercruyssen, F.~D.
Tichelaar, D.~N. Loudkov, and T.~M. Klapwijk,\\
 Kavli Institute of Nanoscience, Faculty of Applied Sciences, Delft
University of Technology, Lorentzweg 1, 2628 CJ Delft, The
Netherlands}

%\maketitle

\begin{abstract}

High critical current-density (10 to 420 kA/cm$^2$)
superconductor-insulator-superconductor tunnel junctions with
aluminium nitride barriers have been realized  using a remote
nitrogen plasma from an inductively coupled plasma source operated
in a pressure range of $10^{-3}$ to $ 10^{-1}$ mbar. We find a
much better reproducibility and control compared to previous work.
From the current-voltage characteristics and cross-sectional TEM
images it is inferred that, compared to the commonly used AlO$_x$
barriers, the poly-crystalline AlN barriers are much more uniform
in transmissivity, leading to a better quality at high critical
current-densities.

\end{abstract}

\maketitle

Quantum technology based on superconducting or magnetic metals
uses nanometer thick tunnel barriers. Many routes to quantum
computation are based on aluminium with aluminium oxide barriers.
Niobium devices use a proximitized layer of aluminium with a
similar oxide \cite{Gurvitch}. Magnetic tunnel junctions have
recently moved from using amorphous aluminium oxide to epitaxial
magnesium oxide with its unique spin-dependent properties
\cite{Parkin,Yuasa}. In quantum computation the amorphous tunnel
barrier has surfaced as an important source of decoherence leading
to the introduction of an epitaxial aluminium oxide barrier
\cite{OhSimmondsMartinis2005,OhSimmonds2006}. On the other hand
highly transmissive tunnel barriers are urgently needed for
sub-millimeter mixers in order to achieve a high bandwidth
\cite{kawamura}, in electronic refrigeration to maximize the
cooling power \cite{Giazotto} and in high density magnetic memory
devices \cite{Tsunekawa2006}.

It has been demonstrated that a major problem of amorphous AlO$_x$
barriers is that they are laterally inhomogeneous
\cite{RippardBuhrmanPRL2002,LangMartinisRevSciInstr2004}. We take
this into account by using a  distribution of transparencies $T_n$
by writing for the voltage-independent normal conductance:
\begin{equation}
G\propto \sum_{n}A_nT_n
\end{equation}
with $A_n$ a fraction of the total area of the tunnel barrier with
an assumed uniform transmissivity $T_n$. Hence, we do not assume a
universal distribution of transparencies
\cite{SchepBauer,NavehPatelAverin} but one which is related to a
distribution of areas with different transmissivities, resulting
from the technological process. For superconducting
tunneljunctions (SIS) this amounts to a situation analogous to
superconducting quantum point contacts \cite{ScheerDevoret}:
\begin{equation}
I\propto \sum_{n}A_n~j(V, T_n)
\end{equation}
with $I$ the total current and $j(V,T_n)$ the voltage-dependent
current-density per area $A_n$. $j$ contains contributions of
different orders proportional to $T_n$, ${T_n}^2$, $T_n^3$,
etcetera, reflecting multiple Andreev reflections
($j(V,T_n)=j_1(V,T_n)+j_2(V,T_n^2)+j_3(V,T_n^3)+\cdots$). For the
commonly used low current-density tunnelbarriers most $A_n$ have
$T_n$ of the order of $10^{-4}$. Since for SIS junctions first
order tunneling $j_1$ leads to a zero subgap current, the
remaining subgap current is due to the higher order terms ($j_2$,
$j_3$, $\cdots$), which only appear for $T_n\approx 1$.
Non-uniformity, causing the emergence of areas with $T_n\approx
1$, thus leads to excessive subgap currents. Therefore the
critical current density of amorphous aluminium oxide barriers is
limited to 20 kA/cm$^2$ \cite{Miller}. We will demonstrate that
aluminium nitride barriers are superior to aluminium oxide
barriers with respect to barrier uniformity.

In the work reported here a very good reproducibility is realized
by using the afterglow region of a nitrogen plasma from an
inductively coupled plasma source (COPRA) (see for example Weiler
\cite{Weiler}), from CCR technology. The plasma provides the
energy to split the N$_2$ molecules into N radicals, needed for
the growth of AlN. The source is mounted on a vacuum chamber. The
plasma is created in the source and diffuses into the chamber. We
have chosen to work in a range of high pressures (2 $\times$
10$^{-3}$ mbar to 1 $\times$ 10$^{-1}$ mbar), for two reasons.
First, we expect at these higher pressures a larger fraction of
atomic N. Secondly, the ion energies in this regime are as low as
a few eV, which minimizes damage to the barrier-formation. This is
different from recent work, where the plasma process not only
provides the chemically active species but also creates damage by
highly energetic ions \cite{Shiota,Bumble,Wang,Kaul,Iosad}
(although usable routes have been reported \cite{Bumble}). In
addition many other plasma techniques suffer from instabilities,
resulting in a poor process reproducibility.

The devices are fabricated on a 2 inch oxidized silicon or fused
quartz substrate. All metal layers are deposited by magnetron
sputtering in the process chamber of a Kurt Lesker system. First,
a 100 nm Nb monitor layer is deposited, after which a ground plane
pattern is optically defined. Subsequently, a bilayer of 100 nm Nb
and about 7 nm Al is deposited. Without breaking the vacuum, the
substrate is then transferred to a nitridation chamber, where the
Al is exposed to the nitrogen plasma for several minutes,
producing a layer of AlN. The substrate is then again \textit{in
vacuo} transferred to the process chamber, where a top electrode
of 200 nm Nb is deposited. The lateral dimensions of the
multilayer of Nb/Al/AlN/Nb are patterned by lift-off. Junctions
are defined by e-beam lithography with a negative e-beam resist
(SAL-601) layer and reactively ion etched (RIE) with a
SF$_6$/O$_2$ plasma using the AlN as an etch-stop, followed by a
mild anodization (5 V). The junction resist pattern is used as a
self-aligned lift off mask for a dielectric layer of 250 nm
SiO$_2$. A 500 nm Nb/50 nm Au top layer is deposited and Au is
etched with a wet etch in a KI/I$_2$ solution using an optically
defined mask. Finally, using an e-beam defined top wire mask
pattern, the layer of Nb is etched with a SF$_6$/O$_2$ RIE, which
finishes the fabrication process.

\begin{figure}[t!]
   \centering
   \includegraphics[width=0.48\textwidth]{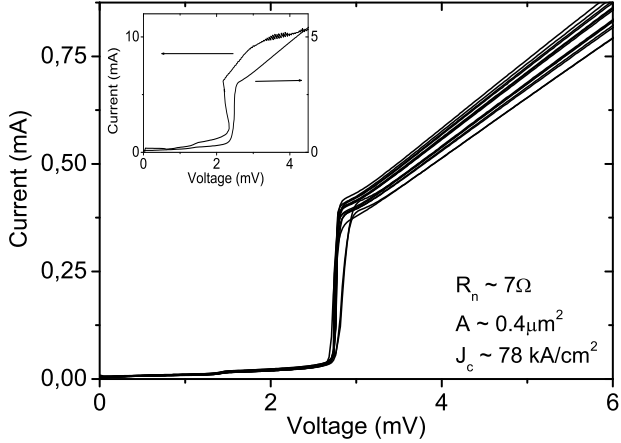}
   \caption{Current-voltage characteristics of a typical batch of
    Nb/AlN/Nb junctions. Junction area is about 0.4 $\mu$m$^2$,
    for a normal resistance, $R_n$, of 6.8 $\Omega$ (critical current density
    $\sim$ 78 kA/cm$^2$). The Josephson current has been suppressed with a
    magnetic field. Inset shows $IV$ characteristics of SIS junctions with
    $J_c$ of 130 kA/cm$^2$ and 420 kA/cm$^2$. For the latter thermal heating
    causes gap-suppression and back-bending.}
\label{iv2605b}
\end{figure}

The fabrication process used provides a very good reproducibility.
There is reproducibility within one fabrication run, illustrated
by the similarity of junctions on a produced wafer. Scatter in the
normal resistance $R_n$ of the junctions is caused by variation in
the junction area $A$, due to uncertainties in junction
definition, and variation in the barrier-specific $R_nA$ value.
For 34 junctions, of which 14 are shown in Fig. \ref{iv2605b}, the
standard deviation,
$\sqrt{\sum_{i=1}^m(R_{n,i}\:-<\!\!R_n\!\!>)^2/(m-1)}$, with $m$
the number of junctions, of $R_n$ has been determined to be 3.3 \%
relative to the average, $<\!\!R_n\!\!>\:=6.8$ $\Omega$. The
peak-to-peak variation amounts to $\pm$6 \%. By measuring four big
junctions (two of 1 $\mu$m$^2$ and two of 2 $\mu$m$^2$), the value
of $R_nA$ has been found to be 2.8 $\Omega\mu$m$^2$ (corresponding
to a critical current density $J_c\approx 78$ kA/cm$^2$). Assuming
perfect junction definition (which is most likely not the case),
the standard deviation of $R_nA$ within one fabrication run is at
most 3.3 \%. Based on the average $R_n$, $A$ is 0.4 $\mu$m$^2$.

\begin{figure}[t!]
   \centering
   \includegraphics[width=0.48\textwidth]{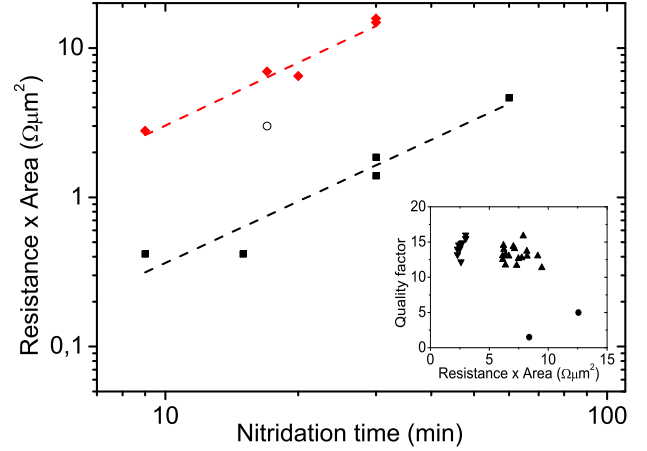}
   \caption{$R_nA$ product as a function of nitridation time, $t_N$,
   for nine different fabricated batches. The squares represent
   data for a 30 cm chuck-source distance, whereas the diamonds
   indicate a 15 cm chuck-source distance. The open circle represents the
   junctions of Fig. \ref{iv2605b}. Dashed lines indicate both
   a dependence $R_nA \propto t_N^{1.4}$. Inset: Quality factor as a function of
   $R_nA$ product for two batches of Nb/AlN/Nb junctions (up- and down-pointing
   triangles). Also indicated
   are AlO$_x$ data from Miller \textit{et al.} \cite{Miller} (filled circles).}
\label{RnA_Ntime}
\end{figure}

We also find a good reproducibility from run to run. We have made
several batches, varying the nitridation time $t_N$ from 9 to 60
minutes. About half of the batches has been made with a low
position of the chuck (30 cm distance to the plasma source) in the
nitridation chamber, the other half with a higher position (15 cm
distance to the plasma source). In Fig. \ref{RnA_Ntime}, we plot
the $R_nA$ product of the batches as a function of $t_N$ for the
large chuck-source distance (squares) and for the small
chuck-source distance (diamonds). The dashed lines indicate a
dependence $R_nA \propto t_N^k$, with $k$ = 1.4. Obviously, there
is a systematic dependence on nitridation time, indicating a
well-behaving process. By varying the nitridation time and/or the
chuck position, we can realize any desired $R_nA$ value between
0.5 $\Omega\mu$m$^2$ and 10 $\Omega\mu$m$^2$.

The quality factor $Q$, defined as $R_{sg}$/$R_n$, where $R_{sg}$
is the subgap resistance, gives an indication of the subgap
leakage through the tunnel barrier. In the inset of Fig.
\ref{RnA_Ntime}, $Q$ has been plotted for two different batches of
AlN based junctions, together with data on AlO$_x$ from Miller
\emph{et al.} \cite{Miller}. In contrast to these AlO$_x$ devices,
it is evident that $Q$ is higher than 10 for all AlN devices. The
lower subgap currents prove that our AlN barriers have a lower
density of areas with $T_n\approx 1$, in other words a better
uniformity.

As shown in Fig. \ref{RnA_Ntime}, we reach $R_nA$ products as low
as ~0.4 $\Omega\mu$m$^2$, corresponding to a $J_c$ of 420
kA/cm$^2$. For such high current densities, heating effects
decrease the superconducting gap voltage of the junction in the
form of back-bending (Fig. \ref{iv2605b} Inset). Up to at least
130 kA/cm$^2$, this effect remains hidden, but is still present.
This indicates that for these AlN barriers the maximum critical
current density is no longer limited by the materials control of
the barrier, but by the intrinsic physical process of
nonequilibrium in the electrodes. Consequently, a future detailed
statistical evaluation of reproducibility and control has to take
into account the distortions of the $IV$ curves by heating.

\begin{figure}[t]
   \centering
   \includegraphics[width=0.48\textwidth]{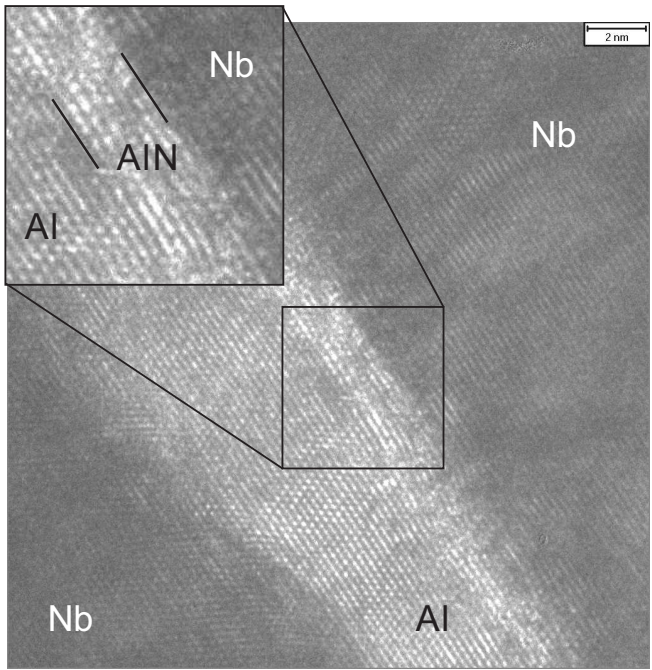}
   \caption{High resolution transmission electron microscope (HRTEM)
   micrographs of an AlN dielectric barrier deposited on an Al layer
   (bright region) between Nb electrodes (dark regions). The bar in the
   top-right corner represents a length of 2 nm.}
\label{TEM}
\end{figure}

Using high-resolution transmission electron microscopy (HRTEM) we
find that the AlN-barrier is grown epitaxially (Fig. \ref{TEM}).
The barrier is visible as a region with higher transmission (most
bright region) and in a small difference in lattice spacing for Al
and AlN. The images clearly indicate epitaxial alignment of the
crystalline structure of the AlN with the underlying Al crystal.
The lattice plane distances of the planes parallel to the surface
were measured in various locations and identified as either
\{0002\} or \{1$\overline{1}$01\} planes in the hexagonal AlN
phase, with spacing 2.49 $\pm$ 0.06 {\AA} and 2.37 $\pm$ 0.06
{\AA} respectively. An averaged thickness of the barrier of about
1.5 $\pm$ 0.5 nm is found. For these devices the $R_{n}A$ value is
about 16 $\Omega\mu$m$^2$. Obviously, in contrast to the commonly
used AlO$_x$, the AlN tunnel barrier has a crystalline nature with
a thickness of about 6 lattice planes, which may be the cause of
the better uniformity.

In conclusion, epitaxial aluminium-nitride tunnel barriers have
been grown, at ambient temperature, using a plasma-source to
generate chemically active nitrogen. This method shows
significantly better reproducibility than other AlN growth
techniques have shown in the past. Compared to the conventionally
used aluminium oxide barriers, much better quality current-voltage
characteristics are observed for high critical current densities,
which is attributed to a spatially more uniform transmissivity of
the epitaxial tunnel barrier.

The authors would like to thank B. de Lange for technical
assistance, and P. C. Snijders and R. W. Simmonds for discussions.
We thank NanoImpuls, Nanofridge, the Dutch Research School for
Astronomy (NOVA), the Dutch Organisation for Scientific Research
(NWO), and the European Southern Observatory (ESO) for funding
this project.

\end{document}